\documentclass{aa}
\usepackage{graphicx}
\usepackage{txfonts}

\usepackage{natbib}

\begin{document}
   \title{A Detailed Observation of a LMC Supernova Remnant DEM~L241
with {\it XMM-Newton}\footnote{%
Based on observations obtained with {\it XMM-Newton},
an ESA science mission with instruments and contributions
directly funded by ESA Member States and NASA.}
}
   \subtitle{}

   \author{Aya Bamba
          \inst{1}
          \and Masaru Ueno\inst{2}
	  \and Hiroshi Nakajima\inst{3}
	  \and Koji Mori\inst{4}
	  \and Katsuji Koyama\inst{3}
          }

   \offprints{Aya Bamba}

   \institute{
     RIKEN (The Institute of Physical and Chemical Research)
     2-1, Hirosawa, Wako, Saitama 351-0198, Japan\\
     \email{bamba@crab.riken.jp}
     \and
     Department of physics, Faculty of Science, Tokyo Institute of Technology
     2-12-1, Oo-okayama, Meguro-ku, Tokyo 152-8551, Japan\\
     \email{masaru@hp.phys.titech.ac.jp}
     \and
     Department of Physics, Graduate School of Science, Kyoto University,
     Sakyo-ku, Kyoto 606-8502, Japan\\
     \email{nakajima@cr.scphys.kyoto-u.ac.jp,koyama@cr.scphys.kyoto-u.ac.jp}
     \and
     Department of Applied Physics, Faculty of Engineering
     University of Miyazaki, 1-1 Gakuen Kibana-dai Nishi
     Miyazaki, 889-2192, Japan\\
     \email{mori@astro.miyazaki-u.ac.jp}
     }

   \date{}

   \abstract{
We report on an {\it XMM-Newton} observation of 
the supernova remnant (SNR) \object{DEM~L241} in the Large Magellanic Cloud.
In the soft band image,
the emission shows an elongated structure, like a killifish,
with a central compact source.
The compact source is point-like,
and named as XMMU~J053559.3$-$673509.
The source spectrum is well reproduced with a power-law model
with a photon index of $\Gamma = 1.57$ (1.51--1.62)
and the intrinsic luminosity is $2.2\times 10^{35}~\mathrm{ergs~s^{-1}}$
in the 0.5--10.0~keV band,
with the assumed distance of 50~kpc.
The source has neither significant coherent pulsations 
in $2.0\times 10^{-3}$~Hz--8.0~Hz, nor time variabilities.
Its luminosity and spectrum suggest that
the source might be a pulsar wind nebula (PWN) in DEM~L241.
The spectral feature classifies this source into rather bright and hard PWN,
which is similar to those in Kes~75 and B0540$-$693.
The elongated diffuse structure can be divided into a ``Head'' and ``Tail'',
and both have soft and line-rich spectra.
Their spectra are well reproduced
by a plane-parallel shock plasma ({\it vpshock}) model with a temperature of
0.3--0.4~keV and over-abundance in O and Ne
and a relative under-abundance in Fe.
Such an abundance pattern and the morphology imply that
the emission is from the ejecta of the SNR,
and that the progenitor of DEM~L241 is a very massive star,
more than 20~M\sun.
This result is also supported by
the existence of the central point source and an OB star association, LH~88.
The total thermal energy and plasma mass are
$\sim 4\times 10^{50}$~ergs and $\sim 200$~M\sun, respectively.
   }

  \keywords{ISM: supernova remnants
    --- X-rays: individual: DEM~L241
    --- X-rays: individual: XMMU~J053559.3$-$673509}

   \maketitle
%

\section{Introduction}

Supernovae (SNe) and supernova remnants (SNRs) shape and enrich
the chemical and dynamical structure
of the interstellar medium and clouds.
X-ray studies give us plenty of information about hot plasma
in SNRs
with emission lines from highly ionized ions.
Moreover, SNRs are believed to be
cosmic ray accelerators
around their pulsar and pulsar wind nebula (PWN),
and/or shock fronts.
The Magellanic Clouds (MCs) are the best galaxies
for the systematic study of SNRs,
thanks to the known distance \citep[50~kpc;][]{feast1999}
and the small absorption column.
Another subject of interest is supernova explosions in starburst galaxies.
The SNe of massive stars scatter light elements into such galaxies,
and make an abundance pattern different from those in normal galaxies
\citep[e.g.,][]{umeda2002}.
The Large Magellanic Cloud (LMC) is the nearest starburst galaxy
\citep{vallenari1996},
and thus we can examine the influence of discrete massive SNe on the galaxy.
Now, more than 30 LMC SNRs are cataloged \citep{williams1999}
and the number may still increase \citep[e.g.,][]{chu2004}.
However,
we have only few samples for which type and age are known.
The pulsars and their nebulae are a clear evidence of core-collapsed origin
and are good indicators of their age.
Now, only four LMC SNRs are reported to have a pulsar and/or PWN;
\object{B0540$-$693} \citep{manchester1993},
\object{N157B} \citep{wang2001},
\object{B0532$-$710} \citep{klinger2002,williams2005},
and \object{B0453$-$685} \citep{gaensler2003}.
This is mainly because of a lack of spatial resolution
in previous radio and X-ray observations.
Thus, searching for new SNR containing pulsars and PWNe in the LMC is
critical to carrying out a systematic study of the contributions from SNRs
to interstellar medium in the LMC.


DEM~L241 (0536$-$67.6) was identified as an SNR
by \citet{mathewson1985}.
They found that
DEM~L241 shows the typical looped filamentary structure of an SNR
with a size of $\sim 2\arcmin~$
in the optical image.
The [\ion{S}{ii}] to H$\alpha$ ratio is 0.6 \citep{mathewson1985},
which is typical of that for SNRs.
Radio emission in 843~MHz was also reported by \citet{mathewson1985},
which shows shell-like structure with similar size to the optical shell.
{\it Einstein} detected relatively bright X-rays
\citep[$2.7\times 10^{35}~\mathrm{ergs~s^{-1}}$ in the 0.15--4.5~keV band,][]{mathewson1985}.
It appears to be a blow-out of the dense \ion{H}{ii} region,
N~59B, around the OB star association \object{LH~88} \citep{chu1988}.
A study of the echelle spectrum by \citet{chu1997} shows
a relatively small expansion velocity and a large intrinsic velocity width,
implying that the SNR may be expanding within the stellar-wind blown cavity,
which suggests that the progenitor of DEM~L241 is likely to be a massive star.
\citet{nishiuchi2001} observed this SNR with {\it ASCA},
and found that
the spectrum of the remnant requires not only thermal emission
but also a power-law component with $\Gamma\sim 1.6$.
This result indicates that
DEM~L241 has a pulsar and/or a PWN,
or synchrotron X-ray emitting shells
like SN~1006 \citep{koyama1995,bamba2003}.
However,
the lack of spacial resolution of {\it ASCA} prevented us from 
concluding the origin of the hard X-ray emission.

In this paper,
we report the detailed X-ray analysis of DEM~L241
for the first time using {\it XMM-Newton},
with the help of {\it ASCA} data.
\S\ref{sec:obs} summarizes
the details of {\it XMM-Newton} and {\it ASCA} observations of DEM~L241.
The analysis results are written in \S\ref{sec:results}.
\S\ref{sec:discuss} is devoted to the discussion about
the origin of the remnant.
We assume the distance to the LMC to be 50~kpc
\citep{feast1999} in this paper.

\section{Observations and Data Reduction}
\label{sec:obs}

{\it XMM-Newton} \citep{jansen2001} EPIC cameras \citep{struder2001,turner2001}
observed DEM~L241 on December 29 2004
(observation ID = 0205380101).
The on-axis point spread function (PSF) is
5\arcsec~ for MOS CCDs
and 6\arcsec~ for pn,
in full width half maximum.
The medium filter was used for all three EPIC cameras
to block ultraviolet photons \citep{stephan1996,villa1998}.
In this observation, both EPIC MOS and pn cameras were operated
in the full frame mode
providing a time resolution of 2.6~s and 73.4~ms, respectively. 
The data reductions and analysis were made
using Science Analysis System (SAS) software version 6.1.0.
We filtered out background flares for rates in $>$10~keV band
more than 0.35 (1.0) counts~s$^{-1}$ for the MOS (pn) cameras,
resulting in 45 (43) ks of good time intervals. 
In the following analysis,
we use the grade pattern 0--12 events for MOS data
and 0--4 events for pn data,
according to the SAS guide.

For the timing analysis,
we used {\it ASCA} \citep{tanaka1994} GIS
\citep[Gas Imaging Spectrometer,][]{ohashi1996}
screened archive data of DEM~L241,
because GIS has better time resolution than pn.
The observation was carried out on October 12--14 1999.
Two GISs (GIS2 and GIS3) were operated in the nominal pulse height mode.
Only high bit data was used,
which provides the best time resolution of 62.5~ms.
The total exposure time is 42~ks for both GISs.
To increase the statistics, the data of the two detectors,
GIS2 and GIS3, were combined in the following study.

\section{Results}
\label{sec:results}

\subsection{Image Analysis}
\label{subsec:image}

\begin{figure}[h]
\centering
\includegraphics[width=8cm]{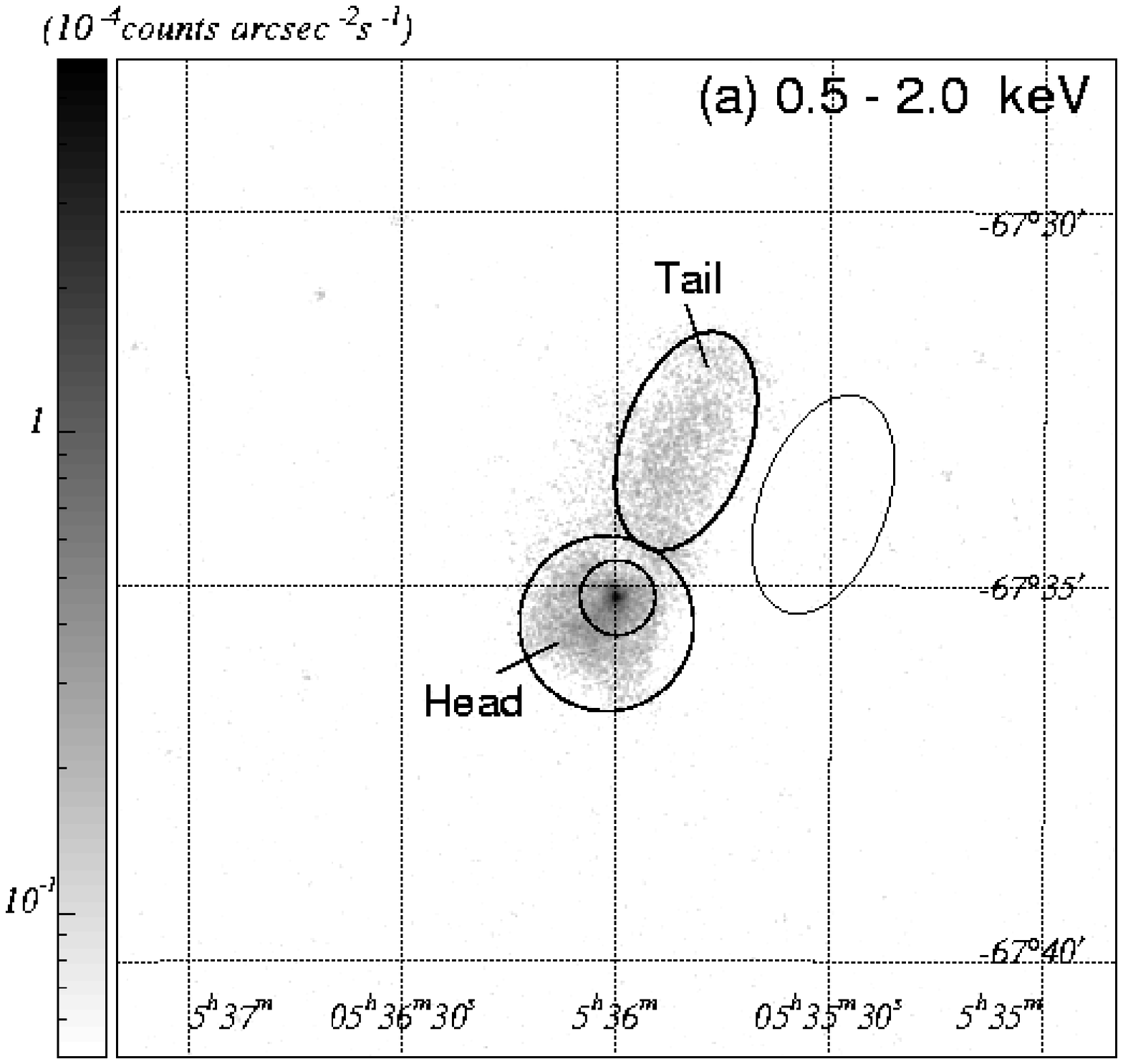}
\includegraphics[width=8cm]{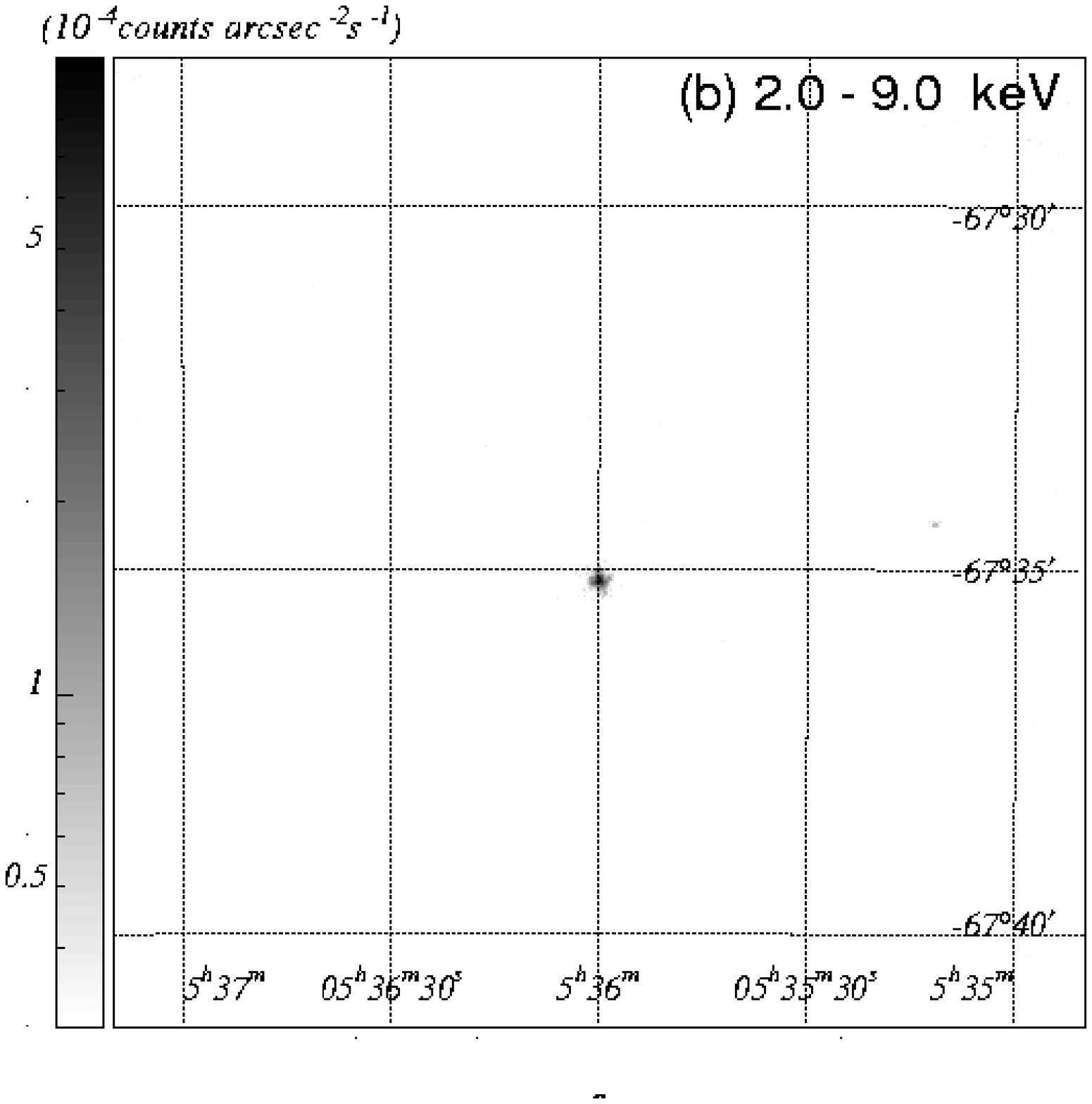}
\caption{MOS 1+2 images in the (a) 0.5--2.0~keV  and (b) 2.0--9.0~keV bands,
with J2000 coordinates.
The scales are logarithmic,
with units of $\times 10^{-4}~\mathrm{counts~s^{-1}arcsec^{-2}cm^{-2}}$,
as shown in the left bar for each image.
Each picture is binned with 20~pixels each and smoothed with 1\arcsec~ scale.
The source and background regions used for the spectral analyses
are shown by the solid thick and thin lines in the panel (a).
}
\label{fig:images}
\end{figure}

Figure~\ref{fig:images} shows the {\it XMM-Newton} MOS~1+2 images of DEM~L241
in the (a) 0.5--2.0~keV and (b) 2.0--9.0~keV bands.
The correction of the exposure time was performed,
whereas the subtraction of background photons was not carried out.
In the soft band image,
we can see a diffuse structure elongated from southeast to northwest
with the size of $\sim 1\farcm5\times 3\arcmin~$,
corresponding to 22~pc$\times$44~pc at 50~kpc.
The shape is like a killifish,
with a double peaked feature on its ``Head'' and ``Tail''
(see Figure~\ref{fig:images}).
In addition to the body of the fish,
there is a compact source like an ``eye'' of the fish.
On the other hand,
only the eye can be seen in the hard band image.

\begin{figure}[h]
\centering
\includegraphics[width=8cm]{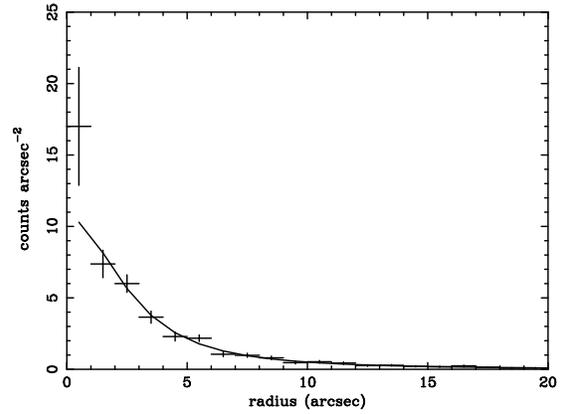}
\caption{MOS~1+2 radial profile of Source~1 in the 2.0--9.0~keV band.
The solid curve and crosses represent the best-fit King profile model and data,
respectively.}
\label{fig:src1_pror}
\end{figure}

We determined the position of the point-like source
with the MOS~1+2 image,
using the {\tt edetect} command in SAS,
to be (05\fh35\fm59\fs93, -67\fd35\fm09\fs72),
then named the source as XMMU~J053559.3$-$673509.
Although counterparts were searched for with the SIMBAD data base,
we found no candidate in any wavelength.
Hereafter, we refer to this source as Source~1.
The radial profile of Source~1 was made
with the 2.0--9.0~keV MOS 1+2 data
as shown in Figure~\ref{fig:src1_pror}.
The source is as compact as the PSF of {\it XMM-Newton}.
The fitting of the profile with a single King profile model was accepted
statistically
with reduced $\chi^2$ of 18.0/17.
The best-fit core radius is 3$\farcs$1 (2$\farcs$4--3$\farcs$9)
(hereafter, the parentheses indicate single parameter 90\% confidence regions),
which is same as the PSF of MOS CCDs.
Therefore, we concluded that
Source~1 is not significantly extended.
The upper-limit of the source size is 1.0~pc.

\begin{figure}[h]
\centering
\includegraphics[width=5cm]{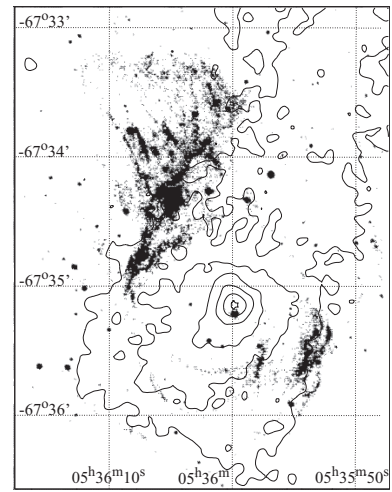}
\caption{The [\ion{S}{ii}] gray-scale map \citep{mathewson1985},
overlaid on the MOS~1+2 0.5--9.0~keV contour map with J2000 coordinates.
The contour is in the logarithmic scale.}
\label{fig:with_SII}
\end{figure}

Figure~\ref{fig:with_SII} shows the [\ion{S}{ii}] gray scale image
\citep{mathewson1985},
overlaid on the MOS~1+2 0.5--9.0~keV contour map.
The [\ion{S}{ii}] emission,
which traces the shock front,
surrounds the Head region
with two strong and weak rims on northeastern and southwestern sides.
The 843~MHz map by \citet{mathewson1985} basically agrees
with the [\ion{S}{ii}] map.
The northern part of the Head is also bright in the light of [\ion{O}{iii}]
\citep{mathewson1985}.
There is no enhancement around the Tail region
in both [\ion{S}{ii}] and radio maps,
although the field of view of the [\ion{S}{ii}] map is too small
to cover the whole Tail region.

\subsection{Spectral Analysis}

\subsubsection{Source~1}
\label{subsec:spec_source1}

\begin{figure}[h]
\centering
\includegraphics[width=8cm]{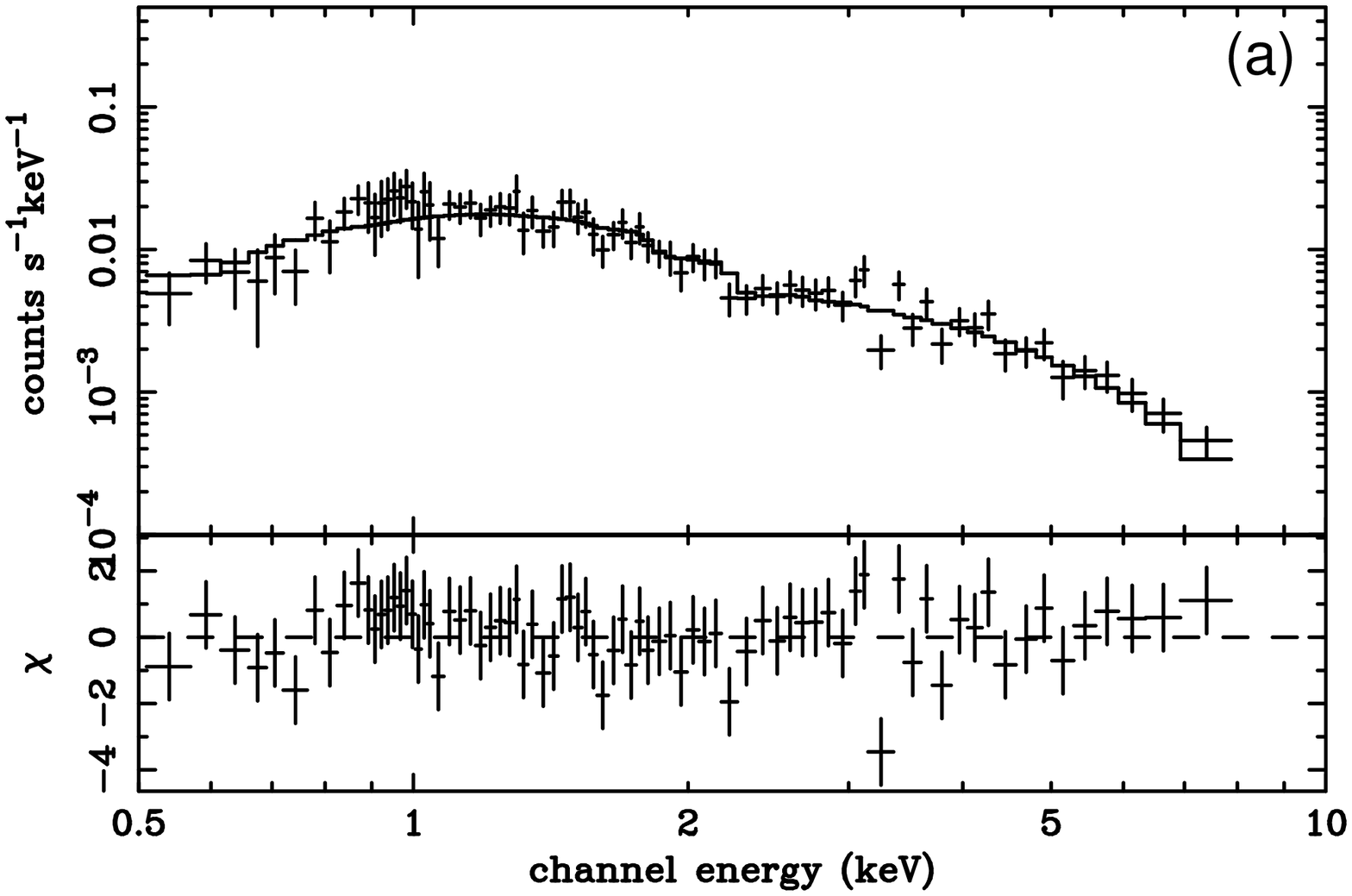}
\includegraphics[width=8cm]{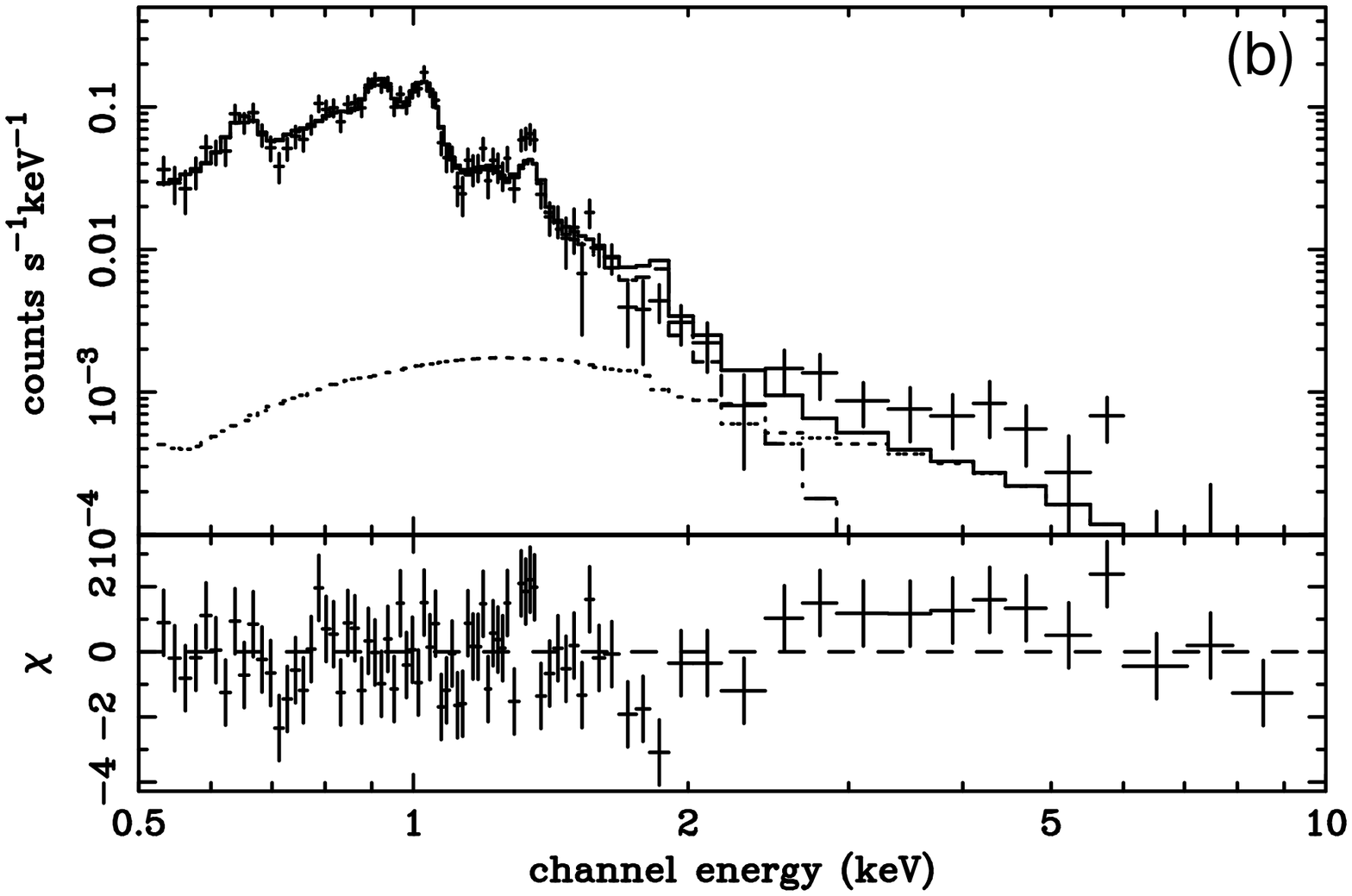}
\includegraphics[width=8cm]{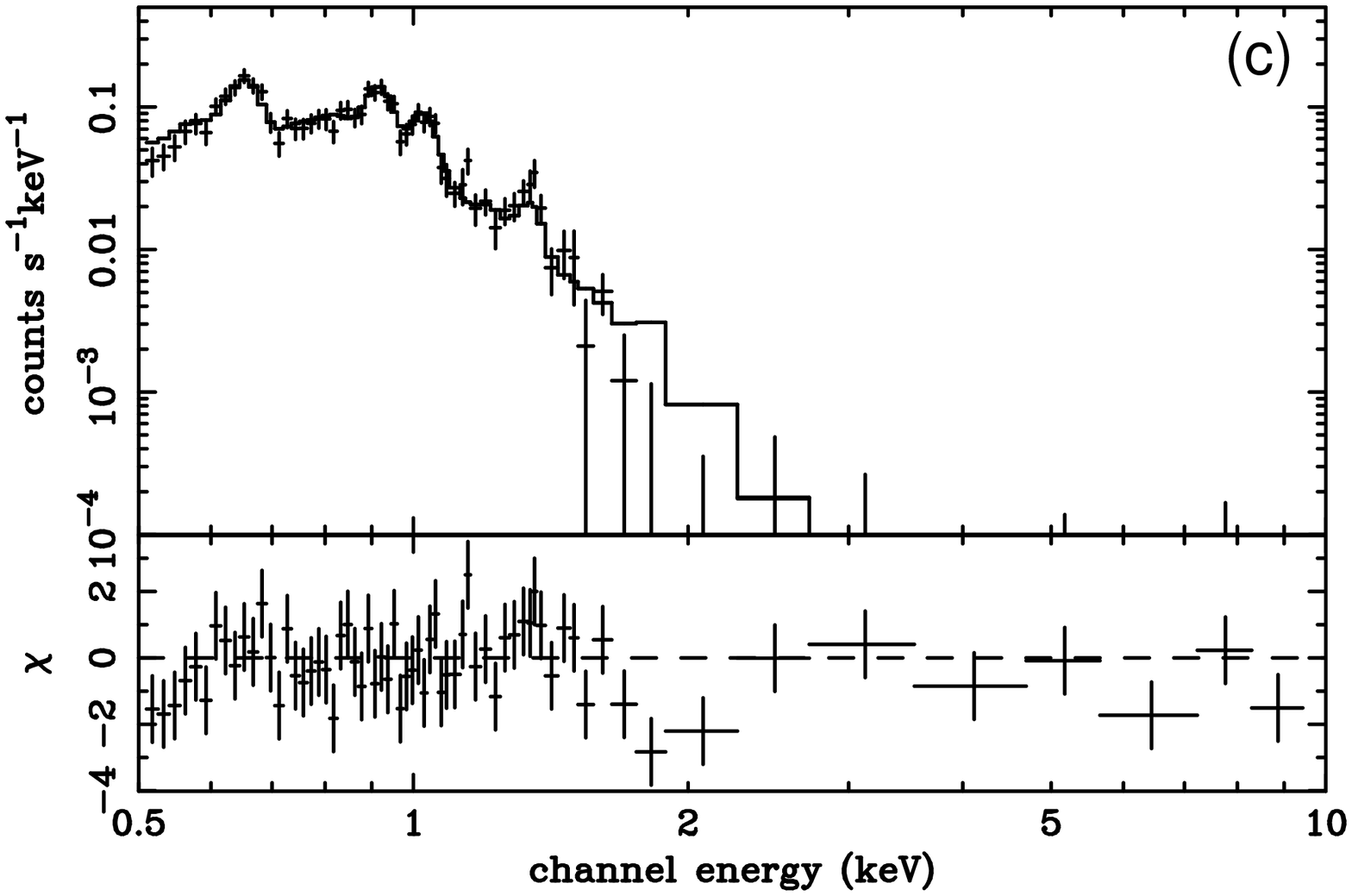}
\caption{Spectra of Source~1 (a), Head (b), and Tail (c) regions.
Although we fitted the spectra of MOS~1+2 and pn CCDs simultaneously,
only the MOS~1 data and results are shown for brevity.
The best-fit models are shown with solid ({\it vpshock})
and dotted (power-law) lines.
Lower panels in figures represent data residuals from the best-fit models.}
\label{fig:spectra}
\end{figure}

The spectrum of Source~1 was accumulated from 
a 15\arcsec~ radius circle around the source for each detector.
The background annular region was selected
with 20\arcsec~ inner radius and 30\arcsec~ outer radius.
Figure~\ref{fig:spectra}(a) shows the background subtracted spectrum
of Source~1,
which is very hard and has no line-like structure.
We fitted the spectra of the three detectors simultaneously
with a power-law function plus absorptions.
The galactic and LMC absorptions
($\mathrm{N_H{}^{gal}}$ and $\mathrm{N_H{}^{LMC}}$, respectively)
were calculated separately.
We fixed $\mathrm{N_H{}^{gal}}$ to be $5.56\times 10^{20}~\mathrm{cm^{-2}}$,
which is estimated by \citet{dickey1990},
whereas $\mathrm{N_H{}^{LMC}}$ was treated as a free parameter.
Each absorption column was subsequently calculated
using the cross sections by
\citet{morrison1983} with the solar abundances \citep{anders1989}
for $\mathrm{N_H{}^{gal}}$,
and the cross sections by \citet{balucinska1992} with
the average LMC abundances \citep[0.3;][]{russell1992,hughes1998}
for $\mathrm{N_H{}^{LMC}}$.
The fitting was acceptable with the reduced $\chi^2$ of 365.1/376.
The best-fit model and parameters are shown in 
Fig.~\ref{fig:spectra}(a) and Table~\ref{tab:spec_src1}.

\subsubsection{Diffuse emission}

The diffuse emission in DEM~L241 shows an elongated structure
with a ``Head'' and ``Tail''.
We divided the emission into the Head and Tail regions
as shown in Figure~\ref{fig:images}(a).
The source region for the spectral analysis of the Head was made from
a 70\arcsec~ radius circle region
excluding a 30\arcsec~ radius circle around Source~1.
On the other hand, the region for the Tail was taken as an elliptical shape
with $50\arcsec\times 90\arcsec$ radii.
The background photons were accumulated from a source free region
near the SNR.
These regions are shown in Figure~\ref{fig:images}(a),
with thick (for the sources) and thin (for the background) lines.

Figure~\ref{fig:spectra}(b) and (c) show the background-subtracted spectra
of the Head and Tail regions.
The spectra are basically soft and have line-like structure,
implying that there is at least a thermal emission component.
The center energies of these lines agree with
He-like O, He-like Ne, and He-like Mg. 
The Head spectrum has an additional hard tail,
which is probably contamination from the bright Source~1,
since the fractional encircled energy within 30\arcsec~ is
only $\sim$90\%.

We used the spectra of the three CCDs simultaneously
and fitted them with 
a plane-parallel shock plasma ({\it vpshock}) model
\citep{borkowski2001},
packaged in {\tt xspec} 11.3.1.,
with the galactic and LMC absorptions.
The absorption calculation has been done in the same way to Source~1 case
(see \S\ref{subsec:spec_source1}).
The abundances of the plasma were fixed to be 0.3
\citep{russell1992,hughes1998}.
For the spectrum of the Head region,
we added an additional power-law component
which represents contamination from Source~1.
The value of $\Gamma$ was fixed to be 1.57, 
the best-fit value for Source~1 (Table~\ref{tab:spec_src1}).
The fittings were rejected with the $\chi^2$/d.o.f. of
871.8/381 (Head) and 607.1/300 (Tail),
with sinusoidal residuals around O and Ne lines.
Therefore,
we allowed O, Ne, and Fe abundances to vary freely
following the previous result by \citet{hughes1998}.
The fits were greatly improved,
and accepted statistically for both regions
with reduced $\chi^2$ of 505.3/377 (Head) and 333.9/296 (Tail).
The best-fit models and parameters are shown in 
Figure~\ref{fig:spectra} (b) and (c),
and Table~\ref{tab:spec_diffuse}, respectively.
We tried to but gave up being free the abundance of Si,
because there is a strong background emission line
\citep{katayama2004}.

The flux of power-law component in the Head region is 
$6.4\pm1.3$\%,
which is consistent with the expected contamination of Source~1.
Thus, 
we concluded that
this hard excess is due to the contamination of Source~1 only,
and do not mention this component hereafter.

\subsection{Timing Analysis}

\begin{figure}[h]
\centering
\includegraphics[width=8cm]{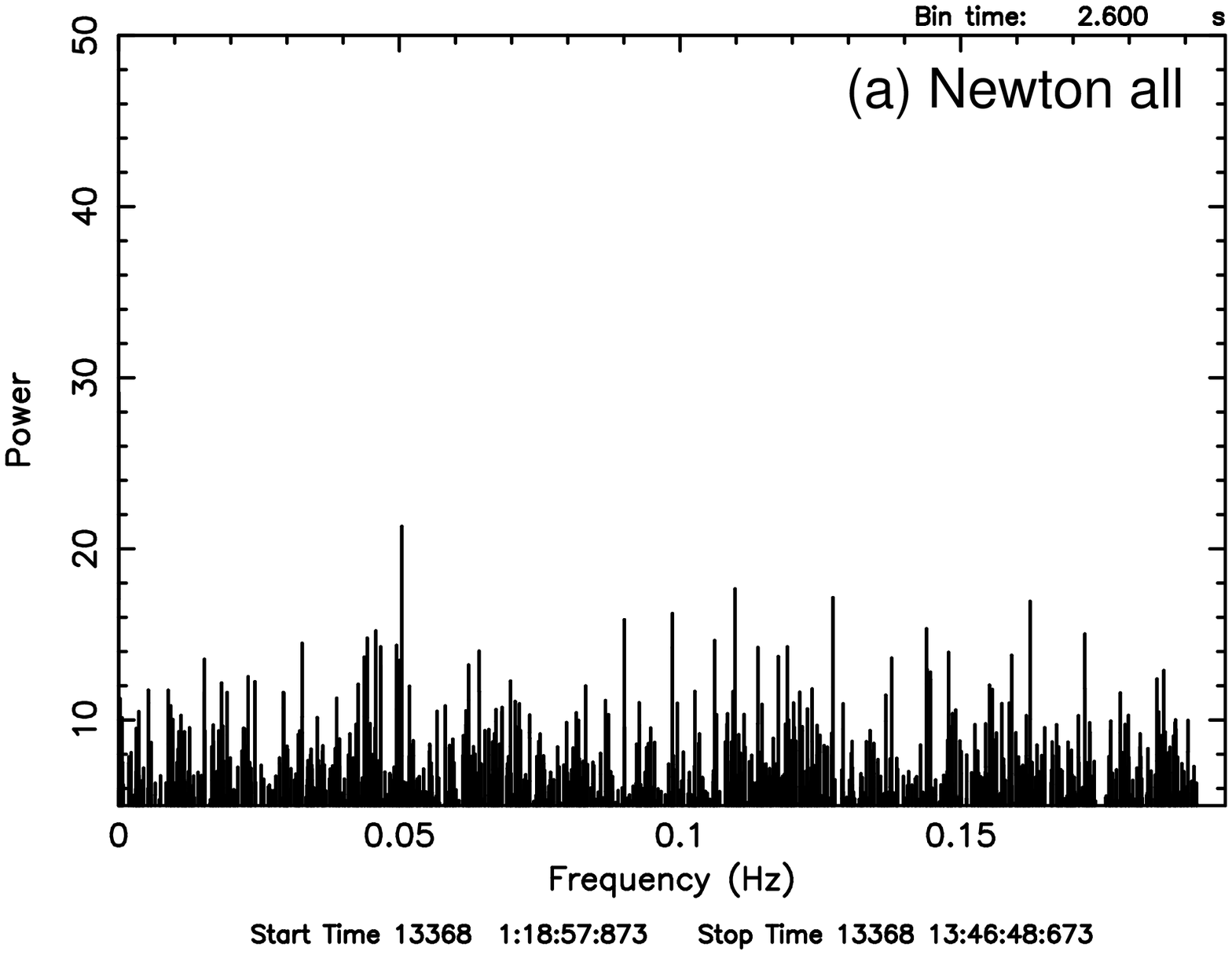}
\includegraphics[width=8cm]{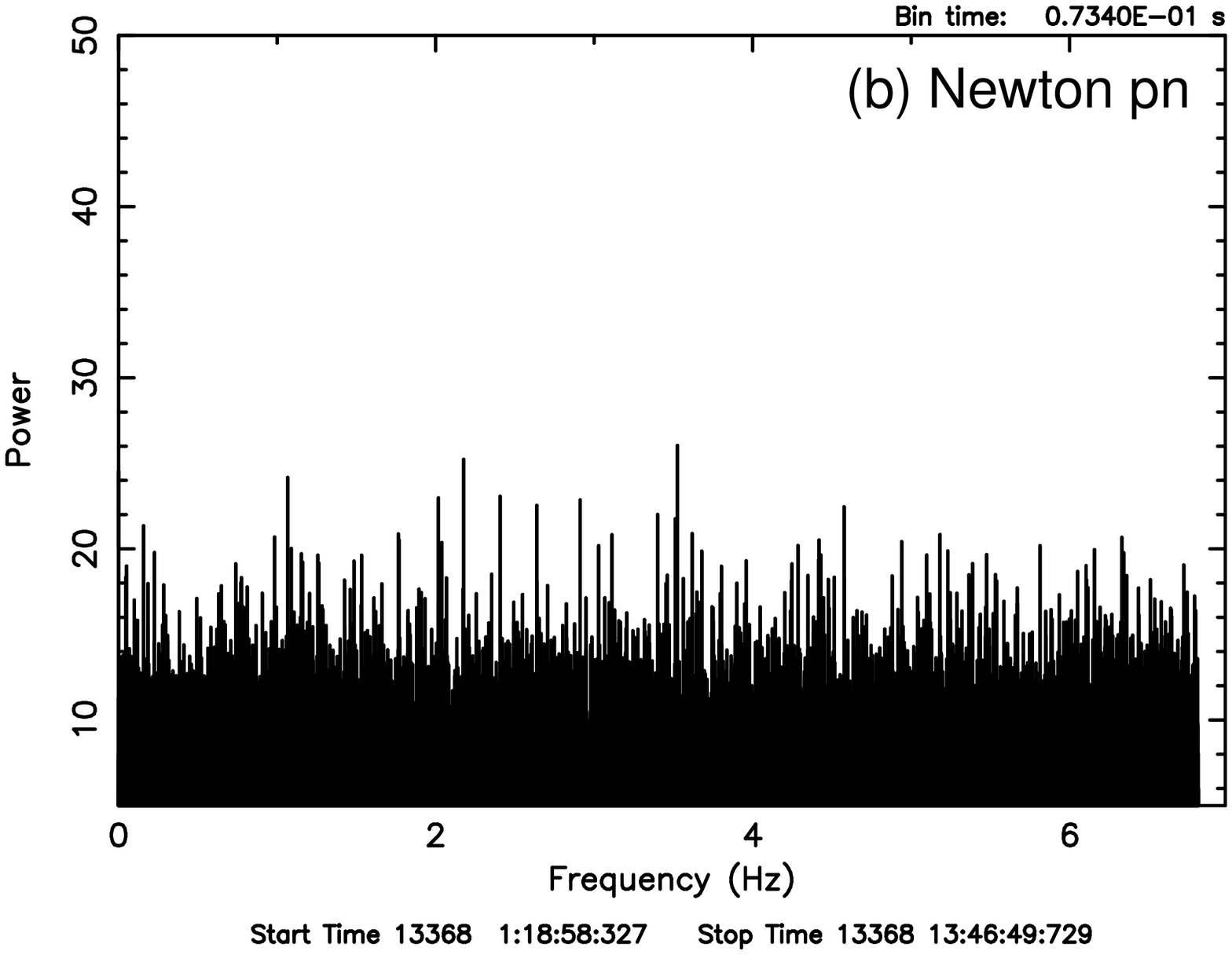}
\includegraphics[width=8cm]{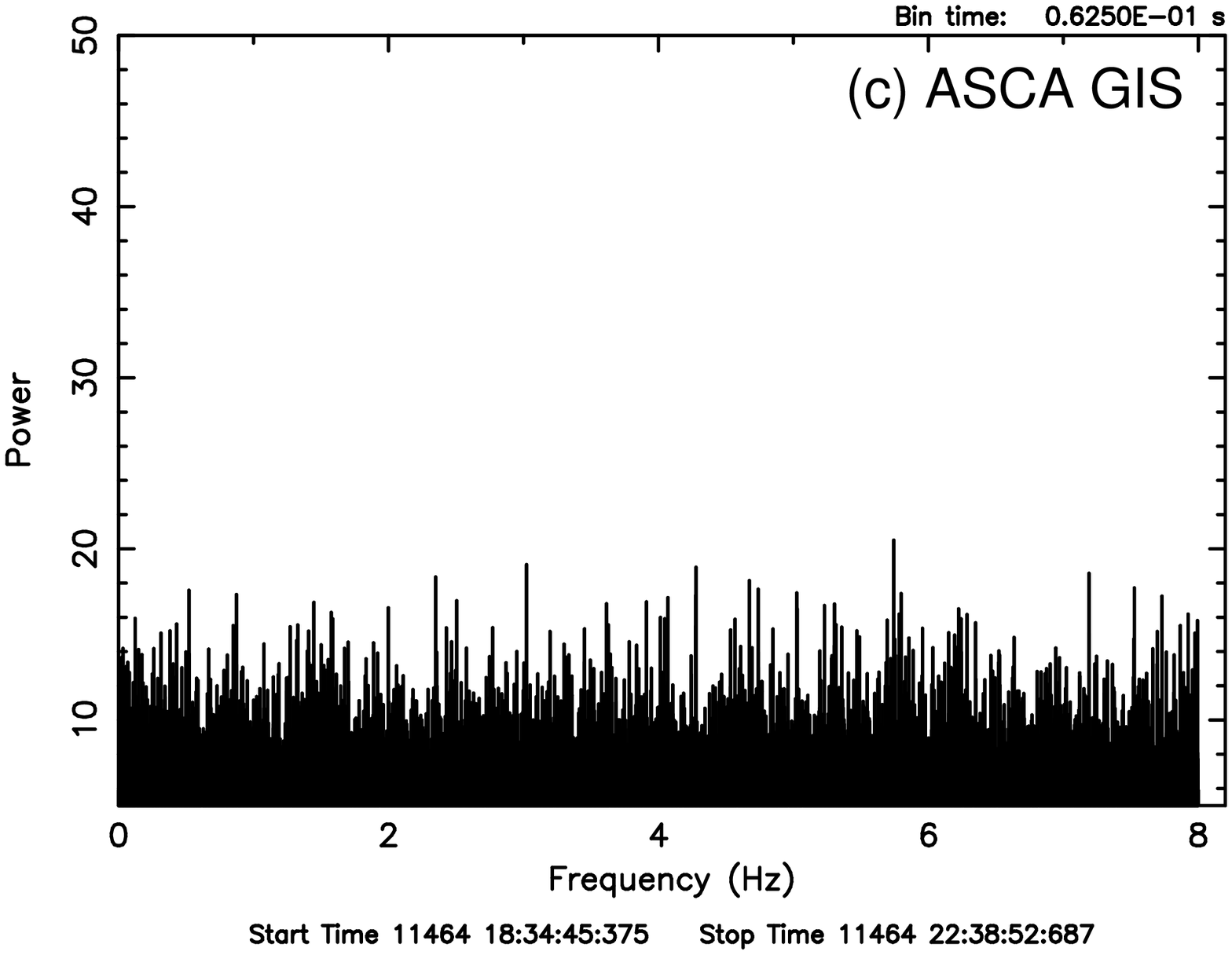}
\caption{The power density spectra
with the MOS1+2 and pn (a), pn (b), and {\it ASCA} GIS2+3 (c).
Data points are plotted only when whose power is larger than 5.}
\label{fig:powspec}
\end{figure}

Coherent pulsations were searched for in Source~1.
We extracted photons from the 2.0--8.0~keV band
(Source~1 is bright in this band as seen in Figure~\ref{fig:spectra}(a))
and from a $15\arcsec~$ radius circle around Source~1.
The correction of the photon arrival times was done
to those at the barycenter.
A Fast Fourier Transform (FFT) algorithm was applied to the pulsation search.
We used all three instruments for the pulsation search
from $2\times 10^{-3}$~Hz to 0.19~Hz,
and only the pn CCD, which has better time resolution than the MOS CCDs,
for the 0.19--6.8~Hz search.
The resulting power density spectra (Figure~\ref{fig:powspec}a, b)
show no significant peak from $2\times 10^{-3}$~Hz to 6.8~Hz.
Although we also analyzed timing data
accumulated from a $5\arcsec$~radius circle
and/or photons in other wider and narrower bands,
the results did not change significantly.

We also conducted the timing analysis with {\it ASCA} GIS data,
which has better time resolution than MOS and pn.
The source photons were accumulated
from a 3\arcmin~ radius circle around Source~1,
within the 2.0--8.0~keV band.
After correcting the photon arrival times to 
those at the barycenter,
we applied a FFT algorithm to search the coherent pulsation.
No significant pulsation was found up to 8.0~Hz again,
as can be seen in Figure~\ref{fig:powspec}(c).

\begin{figure}[h]
\centering
\includegraphics[width=8cm]{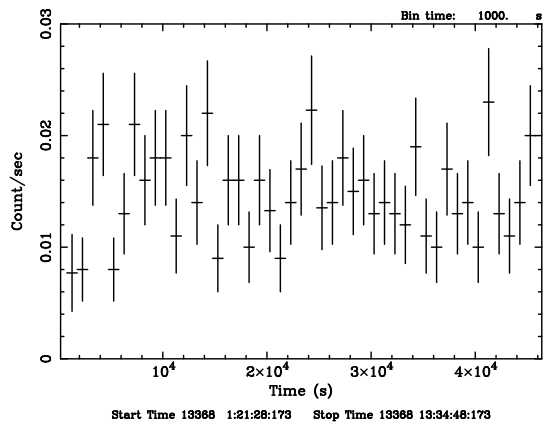}
\caption{X-ray light curves of Source~1 with MOS1+2 and pn CCDs
in the 2.0--8.0~keV band (bin time = 1ks).}
\label{fig:src1_lc}
\end{figure}

Figure~\ref{fig:src1_lc} shows
the light curve of Source~1 in the 2.0--8.0~keV band,
which is the combined result of the three {\it XMM-Newton} detectors.
We have done the Kolmogorov-Smirnov test for this light curve 
and found that Source~1 has no significant time variability
during the observation
with the probability of constancy of 0.72.

\section{Discussion}
\label{sec:discuss}

\subsection{The absorption}

The best-fit $\mathrm{N_H{}^{LMC}}$ in our analysis is
almost consistent with or slightly larger than
the LMC absorption column estimated from the 21~cm line survey
\citep[$8.4\times 10^{20}~\mathrm{cm^{-2}}$;][]{rohlfs1984},
implying that this source is in the LMC.
The small excess might be due to the \ion{H}{i} excess
reported in the \ion{H}{i} map by \citet{stavely-smith1998}.
However, we can conclude nothing
due to the lack of high sensitivity information
around the SNR.
More \ion{H}{i} observations with better spatial resolution
are required to discuss this issue quantitatively.

\subsection{Source~1}

The central position of Source~1, the hard spectrum,
and persistent luminosity
indicate that
Source~1 might be a pulsar and/or a PWN of DEM~L241.
The lack of detection of coherent pulsation is not surprising,
since the periods of ordinary rotation-powered pulsars detected with X-rays
are $\lse$0.1~sec.
The hard spectrum also resembles the emission from 
background active galactic nuclei (AGNs), 
although the source shows no time variability.
We estimate the probability of chance coincidence
that there is a background AGN accidentally
in the SNR region.
With the $\log N$---$\log S$ relation of AGNs
derived by \citet{hasinger1998},
we found that
the expected number of AGNs is only $1.1\times 10^{-3}$
in the $1.5\arcmin\times 3\arcmin$ region.
Therefore, we concluded that Source~1 is an LMC member.
The similarity of the absorption column for Source~1
to that of the Head region also support that.
In the case that Source~1 is a pulsar and/or a PWN,
this adds a new member to the sample of LMC PWNe \citep{gaensler2003}.
The hard X-ray flux detected by {\it ASCA} \citep{nishiuchi2001}
is consistent with that of Source~1.
Thus we consider that
is not nonthermal emission from energetic electrons
accelerated on the shell,
but the central point source.
The distance from Source~1 to the center of LH~88 is about 14\arcsec,
and the diaphragm diameter of LH~88 is 100\arcsec~ \citep{bica1996},
so the progenitor might be a member of the cluster.

The intrinsic luminosity of Source~1 is
$2.2\times 10^{35}~\mathrm{ergs~s^{-1}}$
in the 0.5--10.0~keV band
under the assumption that the distance to Source~1 is 50~kpc,
and the photon index is 1.57 (1.51--1.62).
The luminosity and hardness ranges of pulsars are wide,
$\log L_x \sim $~31--36, $\Gamma \sim$~0.6--2.0, respectively
\citep[e.g.,][]{gotthelf2002,gaensler2003}.
Recently, ``compact central sources'' have been discovered in several SNRs
(Cas~A, \citealt{chakrabarty2001};
Vela~Jr., \citealt{kargaltsev2002};
Kes~79, \citealt{seward2003}),
which have rather small luminosities
($\log L_x \sim $~32--34)
and soft spectra ($\Gamma$ = 3--4).
In the case that 
Source~1 is a pulsar or a compact central source, 
Source~1 is one of the brightest samples,
and as bright as the Crab pulsar \citep{pravdo1997}.
Such bright pulsars certainly have PWNe,
which are about 10~times brighter than the pulsars themselves
and also have hard spectra
\citep[$\log L_x = $33--37 and $\Gamma = $1.3--2.3;][]{gotthelf2002}.
Thus, we conclude that
the emission is from a PWN of DEM~L241.
Its photon index is rather small for ordinary PWNe.
and similar to those of 
the PWN in \object{Kes~75} \citep{helfand2003}
and \object{B0540$-$693} \citep{hirayama2002}.
\citet{helfand2003} notes that
these pulsar systems change their energy into X-ray radiation
very efficiently.
Source~1 may be a new sample of such energetic PWNe.
Source~1 is moderately bright for PWNe,
which might imply that the system is middle-aged.
Pulsars in such PWNe have periods of less than a few hundred msec
and luminosities of about 10\% of their nebulae.
Therefore it is natural that we could not detect the coherent pulsation
even in the case that Source~1 has a pulsar.

In order to confirm the origin of Source~1,
it is essential to resolve the pulsar spatially
and to detect coherent pulsations.
Therefore, further observations are encouraged
with better time and spatial resolution, and excellent statistics,
in the X-ray and radio bands.

\subsection{Diffuse emission}

The temperature and abundance pattern of the hot plasma
in the Head and Tail regions are roughly statistically the same as each other,
(Table~\ref{tab:spec_diffuse}),
implying that these plasmas have the same origin.
The overabundant O and Ne relative to the average LMC
indicate that
the plasma emission is from the ejecta of the explosion,
and that relative to Fe implies that
the progenitor of DEM~L241 may be a core-collapsed SN
\citep{tsujimoto1995}.
The Fe abundance is slightly smaller than
the avarage value in the LMC \citep[0.22][]{hughes1998},
it might be the local fluctuation of metal.
The less abundant O relative to Ne might imply that
the progenitor is very massive, $\gse 20 M$\sun~ \citep{umeda2002}.
\citet{umeda2002} also insists that
fast convective mixing in the progenitor star \citep{sprut1992}
makes more Ne and less O.
Therefore, we concluded that the progenitor of DEM~L241 was
very massive star.
This conclusion is also supported by
the presence of the energetic pulsar candidate
and the OB star association, LH~88 \citep{chu1988}.
The abundance pattern of DEM~L241 resembles in
that of a starburst galaxy \object{M82} \citep{tsuru1997}.
This fact might imply that
DEM~241 have been produced in a period of starburst activity
of the OB star association.
\citet{umeda2000} and \citet{nakamura2001}
have shown that a large Si/O ratio may be signature of 
energetic SN explosion ($\gse 10^{51}$~ergs).
Thus,
we need to determine Si abundance with low background observations.

The thermal emission in the Head region has a center-filled morphology
even considering the contribution from Source~1.
Note that the radius of the extent of the thermal emission is $\sim$70\arcsec~
while that of Source~1 corresponds to the {\it XMM-Newton} PSF,
$\sim$5\arcsec.
The structure,
the radio shell \citep{mathewson1985},
and the large size of the SNR (see \S\ref{subsec:image}),
classify the SNR as a mixed-morphology (MM) SNRs \citep{rho1998}.
However, the over-abundant light elements suggest that
the emission is not from interstellar medium but ejecta
as already mentioned,
which is not common for ordinary MM SNRs.
Then, we concluded that this SNR is an ejecta dominant SNR
without X-ray emitting shells.
The absence of a limb-brightened X-ray shell is expected,
since this SNR is in a hot and tenuous bubble \citep{chu1997}
and it is hard to form the strong shock.
The elongated shape of the Tail region,
together with the lack of enhancement of
[\ion{S}{ii}] and radio continuum emission
around Tail,
might indicate that
the plasma was ejected an-isotropically
into a low-density region.
It is natural to consider that
the plasma in the Tail region is a blown-out
on the basis of the [\ion{S}{ii}] morphology.
\citet{chu1997} suggested that
the SNR could be interacting with the inner walls of a bubble
to produce the SNR signature.
In such a case, the interacting region is heated and emits thermal X-rays,
and the emission certainly traces the shock region,
like the [\ion{S}{ii}] emission.
If the thermal emission is, on the other hand, center-filled,
then we conclude that
the thermal X-rays are not from the interacting region
but from the ejecta.

In order to estimate physical parameters of the plasma,
we assumed that the plasma in the Head region is distributed
in a sphere with the radius of 70\arcsec,
excluding a spherical region with the radius of 30\arcsec~
around Source~1
(total volume $V_{Head} = 5.6\times 10^{59}~\mathrm{cm}^3$),
and that the Tail region has an ellipsoid distribution
with the radii of 50\arcsec$\times$50\arcsec$\times$90\arcsec~
($V_{Tail} = 4.0\times 10^{59}~\mathrm{cm}^3$),
and calculated the mean electron density ($n_e$),
the age of the plasma ($t_p$),
the thermal energy ($E \simeq 3n_eVkT$),
and the total mass of the plasma ($M = n_em_pV$,
where $m_p$ is the proton mass).
The results are summarized in Table~\ref{tab:diffuse_sum}.
The $t_p$ indicates that the SNR is rather aged,
which is consistent with the large size of the SNR.

\section{Summary}

We have conducted a detailed analysis of
DEM~L241 with {\it XMM-Newton} for the first time.
A summary of our results is as follows:
\begin{enumerate}
\item
The X-ray emission shows an elongated morphology
with a Head and Tail like a killifish.
Its size is about 22~pc$\times$44~pc at 50~kpc.
The [\ion{S}{ii}] and radio emission surround the Head,
whereas there is no enhancement around the Tail.
\item
We resolved a central point source,
XMMU~J053559.3$-$673509,
in DEM~L241 for the first time.
The source has no counterpart in any wavelength,
and neither coherent pulsation nor time variability is found.
The spectrum is well reproduced by a power-law model with parameters of
$\Gamma = 1.57$ and
the X-ray luminosity of $2.2\times 10^{35}$~ergs in the 0.5--10.0~keV band,
implying that
this source is a bright hard PWN in DEM~L241.
\item
The Head and Tail have soft and line-rich spectra
with an over-abundance of O and Ne.
The abundance pattern
indicates a core-collapse origin from a massive progenitor
heavier than 20~M\sun~ for DEM~L241.
\end{enumerate}

\begin{acknowledgements}
This research has made use of the SIMBAD database,
operated at CDS, Strasbourg, France.
Our particular thanks are due to R.~M.~Williams,
M.~Nishiuchi, J.S.~Hiraga,
A.~Kubota, I.~Takahashi, K.~Makishima, P.~Ranalli, M.~Nakajima,
and S.~Park,
for their fruitful discussions.
M.~U. and H.~N. are supported by JSPS Research Fellowship for Young Scientists.
This work is supported in part by the Grant-in-Aid for Young Scientists (B) of
the Ministry of Education, Culture, Sports, Science and Technology 
(No.~17740183).
\end{acknowledgements}

\begin{table}[h]
\caption{Best-fit parameters for Source~1$^{\mathrm{a}}$}
\label{tab:spec_src1}
$$
\begin{array}{p{0.45\linewidth}p{0.3\linewidth}}\hline\hline
 \noalign{\smallskip}
Parameters & \\
\hline
$\Gamma$\dotfill & 1.57 (1.51--1.62) \\
$\mathrm{N_H{}^{LMC}}$~[$10^{21}\mathrm{cm}^{-2}$]$^\mathrm{c}$\dotfill & 3.3 (2.7--4.0) \\
Flux~[$10^{-13}\mathrm{ergs~cm^{-2}s^{-1}}$]$^{\mathrm{c}}$\dotfill & 6.4 (6.0--6.8) \\
\hline
\end{array}
$$
\begin{list}{}{}
\item
{$^{\mathrm{a}}$: Parentheses indicate single parameter
90\% confidence regions.}
\item
{$^{\mathrm{b}}$: Calculated using the cross sections by \citet{morrison1983}
with the average LMC abundances \citep{russell1992}.}
\item
{$^{\mathrm{c}}$: In 0.5--10.0~keV band.}
\end{list}
\end{table}

\begin{table}[h]
\caption{Best-fit parameters for the diffuse emission$^{\mathrm{a}}$}
\label{tab:spec_diffuse}
$$
\begin{array}{p{0.4\linewidth}p{0.3\linewidth}p{0.3\linewidth}}\hline\hline
 \noalign{\smallskip}
Parameters & Head & Tail \\
\hline
{\it vpshock}\\
$kT_e$\ [keV]\dotfill & 0.39 (0.38--0.40) & 0.35 (0.34--0.39) \\
$n_et_p$\ [$10^{12}$cm$^{-3}$s]\dotfill & ($>$14.6) & 9.0 ($>$2.0) \\
$[\mathrm{O/H}]$$^{\mathrm{b}}$\dotfill & 0.36 (0.30--0.47) & 0.47 (0.36--0.64) \\
$[\mathrm{Ne/H}]$$^{\mathrm{b}}$\dotfill & 0.82 (0.74--0.96) & 0.79 (0.63--1.0) \\
$[\mathrm{Fe/H}]$$^{\mathrm{b}}$\dotfill & 0.09 (0.07--0.11) & 0.11 (0.09--0.15) \\
$E.M.$\ [$10^{58}$cm$^{-3}$]$^{\mathrm{c}}$\dotfill & 4.4 (3.8--4.8) & 2.8 (2.1--3.5) \\
$\mathrm{N_H{}^{LMC}}$~[$10^{21}\mathrm{cm^{-2}}$]$^{\mathrm{d}}$\dotfill & 5.6 (4.7--6.4) & 1.8 (1.0--3.1) \\
Flux~[$10^{-13}\mathrm{ergs~cm}^{-2}\mathrm{s}^{-1}$]$^{\mathrm{e}}$\dotfill & 3.2 & 3.6 \\
{\it Power-law}$^{\mathrm{f}}$\\
Flux~[$10^{-14}\mathrm{ergs~cm}^{-2}\mathrm{s}^{-1}$]$^{\mathrm{e}}$\dotfill & 4.6 (3.6--5.6) & --- \\
\hline
\end{array}
$$
\begin{list}{}{}
\item
{$^{\mathrm{a}}$: Parentheses indicate single parameter
90\% confidence regions.}
\item
{$^{\mathrm{b}}$: Abundance ratio relative to the solar value
\citep{anders1989}.}
\item
{$^{\mathrm{c}}$: $E.M. = n_e{}^2V$, 
where $n_e$ and $V$ are the electron density and the volume, respectively.}
\item
{$^{\mathrm{d}}$: Calculated using the photometric absorption cross-sections
by \citet{balucinska1992}
with the LMC abundances \citep{russell1992}.}
\item
{$^{\mathrm{e}}$: In the 0.5--10.0~keV band.}
\item
{$^{\mathrm{f}}$: $\Gamma$ is fixed to the best-fit value for Source~1
(1.57).}
\end{list}
\end{table}

\begin{table}[h]
\caption{Physical parameters of the plasma in Head and Tail$^{\mathrm{a}}$}
\label{tab:diffuse_sum}
$$
\begin{array}{p{0.3\linewidth}p{0.3\linewidth}p{0.3\linewidth}}\hline\hline
 \noalign{\smallskip}
Parameters & Head & Tail \\
\hline$n_e$\ [$\mathrm{cm^{-3}}$]\dotfill & 0.28 (0.26--0.29) & 0.27 (0.23--0.30) \\
$t_p$\ [$10^6$yrs]\dotfill & ($>$1.8) & 1.1 ($>$0.2) \\
$E$\ [$10^{50}\mathrm{ergs}$]$^\mathrm{b}$\dotfill & 2.9 (2.2--3.7) & 0.6 (0.5--0.7) \\
$M$\ [$M$\sun]\dotfill & 132 (122--137) & 89 (76--99) \\
\hline
\end{array}
$$
\begin{list}{}{}
\item
{$^{\mathrm{a}}$: Parentheses indicate single parameter
90\% confidence regions.}
\item
{$^{\mathrm{b}}$: Thermal energy $E$ is estimated to be
$3n_eVkT$.}
\end{list}
\end{table}

\end{document}